\documentclass[11pt]{article}
\usepackage{geometry}[a4paper]
\usepackage{lineno}
\usepackage{authblk}
\usepackage{adjustbox}
\usepackage{cite}

\usepackage{mystuff}
\usepackage{hyperref}

\usepackage{dcolumn}
\usepackage{graphicx} 
\pgfplotsset{compat=1.18}

\title{\boldmath Entanglement in typical states of Chern-Simons theory \\}
\author{Charlie Cummings}
\affil{David Rittenhouse Laboratory, University of Pennsylvania,  209 S.33rd Street, Philadelphia, Pennsylvania 19104, USA}
\date{\today}

\begin{document}

\maketitle

\paragraph{Abstract} 
We compute various averages over bulk geometries of quantum states prepared by the Chern-Simons path integral, for any level $k$ and compact simple gauge group $G$. 
We do so by carefully summing over all topologically distinct bulk geometries which have $n$ disjoint boundary tori and a decomposition into space$\times$time of fixed spatial topology.
We find that to leading order in the complexity of the state, the typical state contains many types of multiparty entanglement, proving a conjecture of Balasubramanian et al. 
Additionally, we compute an averaged wave function which captures the leading order statistics of boundary observables in the $n$ torus Chern-Simons Hilbert space. 

\tableofcontents

\section{Introduction} \label{sec:intro}

The possible entanglement structures of many-body systems in quantum field theory is an interesting open problem.
Quantum entanglement of many-body systems behaves much differently than two-party entanglement \cite{walter2017multipartiteentanglement}. 
This is even true for systems with only three subsystems, which can be entangled in two distinct ways up to local unitary transformations.\footnote{More generally, these states fall into different equivalence classes under stochastic local operations and classical communication (SLOCC).}
These two classes of three-body entanglement are referred to as 1) the GHZ type, $|\mathrm{GHZ}\rangle = (|000\rangle + |111\rangle) / \sqrt{2}$, where the density matrix is separable after a partial trace on one qubit, and 2) the W type, $|\mathrm{W}\rangle = (|100\rangle + |010\rangle + |001\rangle)/\sqrt{3}$, where the density matrix remains entangled after a partial trace \cite{walter2017multipartiteentanglement}. Because the dimension of the Hilbert space of $n$ qubits is exponential in $n$, the number entanglement patterns continues to grow rapidly with the number of qubits.

In this paper, we are interested in the patterns of many-body entanglement that can arise in Chern-Simons theory. Chern-Simons theory \cite{chernsimons,Witten:1988hf} is a three-dimensional topological field theory, which means that the wave functions and observables of Chern-Simons theory only depend on the topology of the manifolds they are evaluated on. The topological nature of Chern-Simons theory can be used to encode quantum information in observables that are naturally protected from a wide class of errors that can afflict nontopological quantum computers \cite{Kitaev_2003,freedman2002topologicalquantumcomputation,Freedman2000AMF,Nayak_2008,Melnikov_2018}. This makes the study of entanglement in topological field theories interesting in its own right \cite{Kitaev_2003,freedman2002topologicalquantumcomputation,Freedman2000AMF,Nayak_2008,Melnikov_2018,Nezami_2020,Bao_2022,Saltonswingle_2017,microsoft2025interferometric,PhysRevLett.133.230603,Melnikov_2019,melnikov2024jonespolynomialsmatrixelements,Kitaev_2006, Hosur_2016,Salton_2017,Balasubramanian_2017,Balasubramanian_2018,fiberedlinks}, with Chern-Simons theory as a particular example. 
Furthermore, topological field theories have diffeomorphism invariance at the quantum level \cite{cmp/1104161738}. Thus, modulo the local excitations mediated by a spin-$2$ boson, they can serve as a toy model of quantum gravity: in some sense, they can be thought of as ``quantum gravity without the gravitons''. 
In fact, summing over topologies in topological field theories has been argued to exhibit a toy model of ensemble averaging in AdS/CFT \cite{Dymarsky_2025,Banerjee_2022,PhysRevLett.134.151603,Dymarsky:2025agh}.
This analogy can be made precise in the context of three-dimensional quantum gravity, where the Einstein-Hilbert action is equivalent to a pair of $\SL(2,\R)$ Chern-Simons gauge fields \cite{WITTEN198846}.\footnote{Although these theories agree at the level of the action, the measures of Chern-Simons theory and three-dimensional gravity do not agree. See \cite{Collier_2023} for recent progress in this direction.} The utility of entanglement in quantum gravity is essential at this point \cite{Ryu_2006,vanraamsdonk,Bao_2015,Balasubramanian_2014,Hubeny_2018,Czech_2023,Akers_2022,balasubramanian2024signalsmultipartyentanglementholography,Melnikov_2023,melnikov2023connectomesholographicstates}, and so a better understanding of entanglement in Chern-Simons theory may lead to new insights in quantum gravity as well.

In a previous work \cite{fiberedlinks}, we studied the entanglement structure of ``link states,'' states defined by the Chern-Simons path integral on three-manifolds $\mathcal{M}$ whose boundary $\partial \mathcal{M}$ consists of $n$ disjoint tori $T^2$ (possibly linked together within $\mathcal{M}$). Furthermore, we restricted to the case of \emph{fibered} links, which essentially means that $\mathcal{M}$ can be decomposed into space$\times$time with a spatial manifold of a fixed topology. We explain this condition in more detail below.
The entanglement structure of such states is completely determined by a topological invariant called the ``link monodromy'' of $\mathcal{M}$.
Link monodromies fall into one of three disjoint classes: periodic, hyperbolic, or reducible \cite{NTclass}.
In \cite{fiberedlinks}, it was shown that fibered link states with periodic monodromy always have an entanglement structure which can be thought of as a generalization of the three-party GHZ state defined above. Furthermore, in \cite{Balasubramanian_2017,Balasubramanian_2018}, it was conjectured that links with hyperbolic monodromy have a multipartite entanglement structure which generalizes the three-party W-state defined above, which they verified in numerous examples.\footnote{Their definition of W-like is essentially not-GHZ like: we will instead adopt the latter terminology.} This suggests a sharp connection between multiparty entanglement structures in topological field theories (TFTs) and three-dimensional topology, which would be interesting to make more precise. 

In this paper, we will use this method to prove this conjecture holds for all but possibly a measure zero set of hyperbolic link states. We do so by determining the entanglement structure of a suitably generic fibered link state in Chern-Simons theory. Because hyperbolic links can be thought of as ``generic'' in the moduli space of fibered links  \cite{mappingclass,NTclass}, if the entanglement structure of ``typical'' or ``random'' link states can be determined, this will be the same as the entanglement structure of a random hyperbolic link. 
We present our main result in Sec.~\ref{sec:entropy}: 
most states in the boundary Hilbert space $\Ha(T^2)^{\otimes n}$ with a well defined bulk geometry contain multiparty entanglement which is not GHZ-like, at least to leading order in the complexity of the state. 
We emphasize that this is \emph{not} the entanglement entropy between subregions of a \emph{single} boundary torus.
To make an analogy with AdS/CFT, the average entropy we discuss in this paper is \emph{not} analogous to the Ryu-Takayanagi formula \cite{Ryu_2006} in a single-boundary spacetime, but is instead analogous to the entanglement entropy due to wormholes in a many-boundary spacetime. 

The distribution we use to compute these averages will be carefully constructed. This distribution averages over all possible topologies of three-manifolds $\mathcal{M}$ with $n$ toroidal boundaries, subject to the constraint that the boundary tori of $\mathcal{M}$ form a fibered link. In other words, we will restrict to the case where $\mathcal{M}$ has a ``nice enough'' decomposition into space$\times$time, which we explain below. We will also prove that the average wave function over all link states has a well defined form:
\begin{align}
    \langle \overline{\Psi}\rangle = \sum_c p_c\, \ketbra{\lambda_c}
    \,. \label{eqn:firstavgstate}
\end{align}
for a probability distribution $p_c$ and ensemble of states $\{\ket{\lambda_c}\}$ which we will determine below. 
This average wave function efficiently captures the typical statistics of state-independent observables in the boundary Hilbert space $\Ha(T^2)^{\otimes n}$ of Chern-Simons theory, or its appropriate generalization in any other 3D TFT with a dual rational conformal field theory (RCFT). 

The rest of this paper is organized as follows. In Sec.~\ref{sec:linkstates}, we review the background material and formalism of fibered link states developed in \cite{fiberedlinks}. In Sec.~\ref{sec:randomlinks}, we define what we mean by the average over fibered link complements (the relevant geometries we discussed above). In Sec.~\ref{sec:unnormalizedlinks}, we compute the average state for link states with the ``natural'' normalization given by the Chern-Simons path integral. In Sec.~\ref{sec:normalizedlinks}, we show that if we enforce that each link state is normalized so that $\tr(\overline{\Psi})=1$, the average link state is instead given by \eqref{eqn:firstavgstate}. In Sec.~\ref{sec:entropy}, we show that the typical normalized link state contains a rich multiparty entanglement structure across various parititions of its boundary tori. We conclude with a discussion in Sec.~\ref{sec:discission}.



\section{Link states} \label{sec:linkstates}

\subsection{Link states}

Link states in Chern-Simons theory are discussed in detail in \cite{Balasubramanian_2017,Balasubramanian_2018,fiberedlinks}. We will briefly review them here for completeness, and refer the reader to \cite{fiberedlinks} for more details.

In this paper, we will work with three-dimensional Chern-Simons theories with a compact simple gauge group $G$.
Chern-Simons theory \cite{chernsimons,Witten:1988hf} is an example of a TFT, which means that the observables and wave functions computed using the Chern-Simons path integral do not depend on the fine-grained geometric details of said manifold. It can be defined on a closed three-manifold $M$ via the path integral
\begin{align}
    Z[M]& = \int DA \, \exp(i k I_{CS}[M,A])\,,
    \\
   I_{CS}& = \frac{1}{4\pi} \int_M d^3 x  \, \Tr[A \wedge dA + \frac{2}{3} A \wedge A \wedge A]  \,.
\end{align}
Here, $A$ is a gauge field for a compact simple Lie group $G$, and $k$ is an integer-quantized coupling constant known as the level of the theory.\footnote{The Chern-Simons action $I_{CS}$ can be shifted by multiples of $2\pi$ by gauge transformations that are not continuously connected to the identity.  This implies that the level $k$ must be an integer, to ensure that $Z[M]$ is well defined.} 

\begin{figure}
    \centering
    \includegraphics{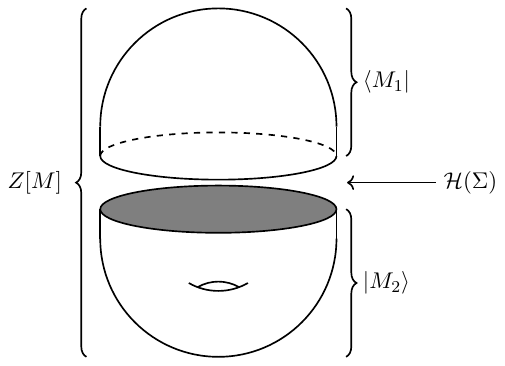}
    \caption{Closed three-manifolds $M$ can be Heegaard split along a two-dimensional submanifold $\Sigma$ as $M = M_1 \sqcup_\Sigma M_2$. The path integral $Z[M]$ computes the inner product of states $\ket{M_1}, \ket{M_2}$, each of which lives in the Hilbert space $\Ha(\Sigma)$. }
    \label{fig:heegaard}
\end{figure}

Let $M$ be a closed three-manifold, and let $\Sigma$ be some closed two-dimensional submanifold  $\Sigma \subset M$. If we split $M = M_1 \sqcup_\Sigma M_2$ along $\Sigma$, we can think of $M$ as the gluing of two manifolds-with-boundary along $\Sigma$.\footnote{This is called a Heegaard splitting of $M$.} The Chern-Simons path integral can be thought of as computing an overlap between states
\begin{align}
    Z[M] = \braket{M_1}{M_2}_\Sigma \,.
\end{align}
The states $\ket{M_1}, \ket{M_2}$ are defined on a Hilbert space $\Ha(\Sigma)$ associated to the splitting surface $\Sigma$, as indicated by Fig.~\ref{fig:heegaard}. Thinking of $\Sigma$ as a ``moment in time,'' $\ket{M_1}$ can be thought of as being prepared with a Hartle-Hawking prescription \cite{hhwavefn} for the path integral $Z[M_1]$ on the manifold-with-boundary $M_1$, with gluing boundary conditions on $\partial M_1 = \Sigma$, and similarly for $\ket{M_2}$. 

As above, let $M$ be a closed three manifold without boundary, and $\La^n$ be an $n$-component link in $M$. A link $\La^n$ is defined to be an embedding
\begin{equation}
    \La^n: \bigsqcup_{i=1}^n S^1 \into M
\end{equation}
of $n$ disjoint circles into $M$. The \emph{link complement} of $\La^n$ in the background $M$ is defined as
\begin{equation}
    M(\La^n) = M \setminus \mathcal{N}(\La^n)\,,
\end{equation}
where $\mathcal{N}(\La^n)$ is a tubular neighborhood of the link in $M$. The picture to keep in mind is that the link complement is constructed by ``drilling out'' $n$ solid tori, linked together in $M$ according to $\La^n$.
Because the only boundary components of $M$ arise from removing these tubular neighborhoods, the boundary of the link complement is
\begin{equation}
    \partial M(\La^n)  = \bigsqcup_{i=1}^n T^2\,,
\end{equation}
i.e., $n$ disjoint copies of the usual 2-torus. From the perspective of the boundary Hilbert space $\Ha\left(\partial M(\La^n)\right)$, the fact that the boundary tori are disconnected means that the Hilbert space on $\partial M$ factorizes: 
\begin{equation}
    \Ha(\partial M(\La^n)) = \bigotimes_{i=1}^n \Ha(T^2)\,.
\end{equation}
Because of this tensor product structure, we can study the entanglement of such states across arbitrary bipartitions of the tori: we are interested in how the different boundary tori are entangled to each other. As we said above, a given link complement $M(\La^n)$ prepares a specific link state $\ket{M(\La^n)}$ in this Hilbert space. If the background manifold is clear from context, we will sometimes write $\ket{\La^n}$ instead.

It turns out that $\Ha(T^2)^{\otimes n}$ is finite dimensional \cite{Witten:1988hf,Moore:1988qv}, and each $\Ha(T^2)$ has a natural basis which is labeled by level-$k$ representations of the gauge group $G$. We will denote natural product basis of $\Ha(T^2)^{\otimes n}$ as 
\begin{equation}
    \ket{J} :=\ket{j_1} \cdots \ket{j_n} \,, \label{eqn:Jbasis}
\end{equation}
where $j_i$ can be thought of as an integrable level-$k$ representation of the gauge group $G$. 

We could imagine defining a similar Hilbert space $\Ha(g,J)$ for a more general surface $\Sigma_{g,J}$ with genus $g$ and $n$ defects, with each defect labeled by representations $j_1, \cdots, j_n$ of the gauge group $G$ at level $k$. The detailed structure of the Hilbert space $\Ha(g,J)$ will not be too important for our purposes. The features which will matter are that 1) each factor $\Ha(g,J)$ is finite dimensional \cite{Witten:1988hf}, and 2) each factor carries a representation of the \emph{mapping class group} $\text{Mod}(g,n)$ \cite{Moore:1988qv}. 
The mapping class group $\text{Mod}(g,n)$ can be thought of as the set of ``large'' diffeomorphisms which maps $\Sigma_{g,J}$ back to itself \cite{mappingclass}. In other words, it is the set of maps $f:\Sigma_{g,J} \to \Sigma_{g,J}$, together with the equivalence relation $f \sim g$ if $f \circ g^{-1}$ is an isotopy of $\Sigma_{g,J}$.\footnote{An isotopy is a diffeomorphism which is continuously connected to the identity, i.e., a ``small diffeomorphism''.} We will denote this equivalence class by $[f]$. Furthermore, we demand that all such equivalence classes $[f]$ have a representative for which there exists an open neighborhood of each defect where $[f]$ acts trivially. Together with the usual composition of maps, these equivalence classes form a group.

\subsection{Fibered links}

Any link $\La^n \subset M$ has a \emph{Seifert surface} $\Sigma_S$, which is a two-dimensional submanifold of $M$ such that $\partial \Sigma_S = \La^n$ \cite{rolfsen1976knots}. For example, the unknot $0_1$ in $S^3$ has a Seifert surface given by a disk spanning the unknot. Such a surface always exists, regardless of the link or background \cite{rolfsen1976knots}. Given a Seifert surface $\Sigma_S$, we can always construct another by attaching additional handles to $\Sigma_S$, raising its genus. Therefore, it is convenient to use a convention where $\Sigma_S$ has the minimal possible genus for a given link and background, which we adopt for the rest of this paper. If a link $\La^n \subset M$ has a unique Seifert surface up to isotopy, then we call $\La^n$ a fibered link, and say that $\La^n$ fibers in $M$. At least in $S^3$, fibered links are not rare; for any nonfibered link $(\La')^n$ in $S^3$, there is a fibered link $\La^{n+1}$ which contains $(\La')^n$ as a sublink \cite{stallings}. In other words, any nonfibered link can be made fibered by introducing a single additional link component.

Because $\Sigma_S$ is a two-dimensional, genus $g$, $n$-boundary Riemann surface, we can identify it (possibly after a homeomorphism) with the standard Riemann surface $\Sigma_{g,n}$ with genus $g$ and $n$ punctures or defects. The difference between $\Sigma_{g,n}$ and $\Sigma_S$ is that $\Sigma_{g,n}$ has contractible boundary components, while $\Sigma_S$ has contractible boundary components only if its link components are unlinked. The effects of this identification are discussed in detail in \cite{fiberedlinks}, and will not be relevant for our purposes. If $\La^n \subset M$ is a link with Seifert surface homeomorphic to $\Sigma_{g,n}$, we call $g$ the genus of the link. 

\begin{figure}
    \centering
    \includegraphics[width=0.5\linewidth]{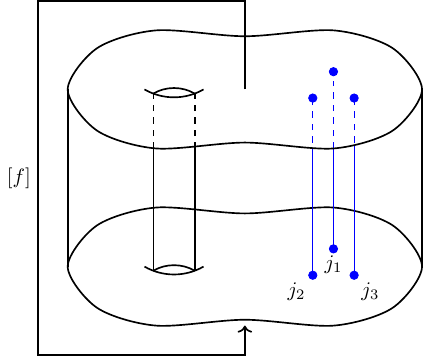}
  
    \caption{Link complement of a fibered link. Here, the Seifert surface of the link is homeomorphic to $\Sigma_{1,3}$. A puncture of $\Sigma_S$ becomes a Wilson loop when traversing the $S^1$ of the fibration, as shown in blue.
    }
    \label{fig:fibration}
\end{figure}

For the rest of this paper, we assume that $\La^n$ fibers in $M$. Now consider deleting a copy of $\Sigma_S$ from the link complement $M(\La^n)$. Because $\Sigma_S$ is unique up to isotopy for fibered links, this implies that
\begin{align}
    M(\La^n)\setminus \Sigma_S \cong \Sigma_S \times [0,1] \,. \label{eqn:splitcomplement}
\end{align}
To reconstruct $M(\La^n)$ from this splitting, we must identify the endpoints of this interval via a map $f:\Sigma_{g,n} \to \Sigma_{g,n}$. The map $f$ glues the two ends of \eqref{eqn:splitcomplement} by the identification
\begin{align}
    \Sigma_S \times \{1\} &\cong f(\Sigma_S) \times \{0\}\,, \label{eqn:gluingwithf}
   \\ M(\La^n) &= \Sigma_{g,n} \times_f S^1 \,. \label{eqn:fibration}
\end{align}
By \eqref{eqn:fibration}, we mean that \eqref{eqn:splitcomplement} has had the identification \eqref{eqn:gluingwithf} applied.
Any other map $f'$ which is isotopic to $f$ will produce the same link complement $M(\La^n)$ after the identification given by \eqref{eqn:gluingwithf} \cite{mappingclass}. Thus, we should think about this map $f$ as being a representative $[f]$ of the mapping class group $\text{Mod}(g,n)$. The mapping class group element $[f]$ is called the monodromy of the link complement $M(\La^n)$. The triple $(g,n,[f])$, a shorthand for \eqref{eqn:fibration}, uniquely determines the link complement $M(\La^n)$ up to homeomorphism. This triple is called an open book decomposition of the link complement \cite{rolfsen1976knots}. See Fig.~\ref{fig:fibration} for an example of a link complement from this perspective.

\subsection{Fibered link states}

Let $M$ be a closed three manifold and $\La^n$ be a link with a fixed genus $g$, $n$ components, and monodromy $[f]$. Such a link state can be written in the (un-normalized) form \cite{fiberedlinks}
\begin{align}
    \ket{M(\La^n)} =  \sum_{J} \tr(K_f[J]) \mathcal{S}\ket{J}\,. \label{eqn:fiberedlinkstate}
\end{align}
Here, $J$ labels an orthonormal basis for the boundary Hilbert space $\Ha(T^2)^{\otimes n}$ as in \eqref{eqn:Jbasis}, $\mathcal{S} = S^{\otimes n}$ is the modular $S$ transformation acting on each boundary torus, and $K_f[J]$ is a unitary representation of $\text{Mod}(g,n)$ which represents the action of the monodromy $[f]$ on the bulk Hilbert space $\Ha(g,J)$.

To briefly explain how this equation arises, the first step is to view the splitting of the link complement in \eqref{eqn:splitcomplement} as defining $\Sigma_S$ as ``spatial'' directions, and the interval $[0,1]$ as a ``time direction''. Then, we fill the boundary tori of $M(\La^n)$ with Wilson lines traversing the $S^1$ of \eqref{eqn:fibration}; quantum mechanically, this corresponds to computing the Chern-Simons path integral\footnote{The modular $S$ transformation arises because the vertical Wilson lines of Fig.~\ref{fig:fibration} are related to the boundary cycles of $\Sigma_S$ by an $S$ transformation.}
\begin{align}
    \bra{M(\La^n)} \mathcal{S} \ket{J} = Z[\Sigma_{g,J} \times_f S^1] \,.\label{eqn:amplitude}
\end{align}
The monodromy $[f]$ can be thought of as the global ``time evolution'' along the $S^1$, which we represent with the operator $K_f[J]$. This operator acts on the genus $g$, defect $J$ Hilbert space $\Ha(g,J)$ associated to the Seifert surface of the link. 
The fact that this ``time evolution'' is enacted along a circle $S^1$ is what leads to the appearance of the trace $\tr(K_f[J])$. Finally, we use a resolution of the identity and \eqref{eqn:amplitude} to compute
\begin{align}
    \ket{M(\La^n)} &= \sum_{J} \mathcal{S} \ketbra{J} \mathcal{S}^\dagger \ket{M(\La^n)} 
    \\&= \sum_{J} Z[\Sigma_{g,J} \times_f S^1] ^* \mathcal{S}\ket{J}\,.
\end{align}
As explained above, we can make the substitution
\begin{align}
    Z[\Sigma_{g,J} \times_f S^1] = \tr(K_f[J])^*\,,
\end{align}
and arrive at \eqref{eqn:fiberedlinkstate}. The complex conjugate is because of our convention for the direction of the monodromy being a map from $\Sigma_S \times \{1\}$ to $\Sigma_S \times \{0\}$. This is the standard convention in the math literature. An equivalent way to represent this pure state is via its un-normalized (pure) density matrix 
\begin{align}
    \Psi&=  \sum_{J,L} \Psi_{J L} \mathcal{S}\ketbra{J}{L} \mathcal{S}^\dagger \,, \label{eqn:matrixoflinkstate}
    \\ \Psi_{JL} &=  \tr(K_f[J])\tr(K_f[L])^*\,.
\end{align}
This is the form of the link state we will focus on for now. We will discuss the effects of normalizing the link state in Sec.~\ref{sec:normalizedlinks}.

\section{Random Links} \label{sec:randomlinks}

\subsection{The definition of a random link}

As explained above, a link complement is uniquely defined up to homeomorphism by its open book decomposition $(g,n,[f])$. We define a random link complement by first fixing $(g,n)$, and then varying the monodromy $[f]$. This definition of random link not only averages over embeddings $\La^n$ of circles into a fixed background $M$, but also sums over the possible background manifolds $M$. This definition of ``random'' link states has various motivations. 

First, we note that that the monodromy of a fibered link is the link invariant which determines the entanglement structure of its associated link state \cite{fiberedlinks}. Thus, averaging over monodromy is a necessary condition to explore the possible entanglement patterns of fibered link states. 

Second, because $K_f$ is a representation of the mapping class group, we can choose a finite set $\mathcal{T}$ of generators for $\text{Mod}(g,n)$ to construct $K_f$,
\begin{align}
    K_f = K_{t_1} \cdots K_{t_N}
\end{align}
for some $t_1, \cdots, t_N \in \mathcal{T}$. If we think of each $K_{t_i}$ as a ``gate'' which is part of the preparation procedure for the link state, then averaging over monodromy is equivalent to averaging over the possible quantum circuits preparing fibered link states. This is the consistent with the view advocated for in \cite{fiberedlinks}, where we view the monodromy as a form of ``time evolution'' for preparing the link state. From this perspective, averaging over monodromy can be thought of as averaging over Hamiltonians preparing the link state. This matches other conventions for random quantum states \cite{Fisher_2023}. 

Finally, a fibered link state is completely determined by its open book decomposition $(g,n,[f])$, as reviewed in Sec.~\ref{sec:linkstates}. So if we averaged a link state over possible monodromy for fixed $(g,n)$, we could then average this result over $(g,n)$ if we wished. This exhausts the possible averaging over the moduli space of fibered link states. As the average over $(g,n)$ is simpler than averaging over monodromy (they are just two positive integers), we will mostly focus on the monodromy average in this paper. 

With this definition of random links, we need to define a sensible way to average over possible monodromies $[f]$ in order to extract the universal behavior of random fibered link states. This poses technical difficulties because of the structure of $\text{Mod}(g,n)$, which is the space we wish to average over. $\text{Mod}(g,n)$ is not a compact group, and so constructing a well defined, uniform average over the entire group will take some work.

\subsection{Defining the average over the mapping class group} \label{sec:measures}

Consider a fixed and finite set $\mathcal{T}$ of generators for the discrete group $\text{Mod}(g,n)$. Because we are interested in multipartite entanglement, we restrict $n \geq 3$.\footnote{Furthermore, to avoid an edge case, we ignore the case $(g,n) = (0,3)$. $\text{Mod}(0,3)$ is isomorphic to the symmetric group $S_3$, and so has no pseudo-Anosov elements \cite{mappingclass}. By the Nielsen-Thurston classification \cite{NTclass}, this means there are no three component hyperbolic links with genus $0$. This is consistent with the examples $K_3$ and $H_3$ of \cite{fiberedlinks}.} By definition, $\mathcal{T}$ generates the mapping class group, so any elements $[f],[g]$ can be expanded as a product of elements of $\mathcal{T}$. For a given group element $[f]$, this expansion is generally not unique, but can always be taken to be finite. Therefore, the \emph{word metric}
\begin{align}
    d([f],[g]) = \min\{N \in \N \,\vert \,\exists \,t_i \in \mathcal{T} \cup\mathcal{T}^{-1}\, \\\text{ such that } [f^{-1} g] =  t_1 \cdots t_N \}
\end{align}
is well defined. The word metric counts the minimal number of generators needed to ``move'' from $[f]$ to $[g]$. One can check that this really does define a metric on $\text{Mod}(g,n)$ \cite{cannon_geometric_group_theory}.
One can also think of the distance $d(e,[f])$ from the identity to $[f]$ as a measure of complexity for $[f]$, as it measures the minimal number of ``gates'' from $\mathcal{T}$ that are needed to construct it \cite{MR1922899}.
This metric depends on a choice of gates $\mathcal{T}$ \cite{cannon_geometric_group_theory}, but these differences will ultimately be unimportant.

In order to average various quantities such as the wave functions of fibered link states over $\text{Mod}(g,n)$, we need to construct the measure we wish to use to compute this average. We will use the word metric to construct a useful family of measures on $\text{Mod}(g,n)$, indexed by $N$, as follows. The measure we are actually interested in will involve a limit of this family. Let $B_N(e)$ be the ball of radius $N$ on $\text{Mod}(g,n)$ with respect to the word metric. In other words, $[f] \in B_N(e)$ if and only if $d(e,[f]) \leq N$. Next, let 
\begin{align}
    F: \text{Mod}(g,n) \to \C
\end{align}
be a bounded function on the mapping class group.\footnote{A bounded function is a function with a constant $K$ such that $|F(x)| \leq K$ for all $x \in \text{Mod}(g,n)$.} We define $\Delta_N$ to be the distribution dual to the indicator function on $B_N(e)$. This means that as a distribution from functions $F$ to the complex numbers $\C$,
\begin{align}
    \Delta_N(F) = \frac{1}{\Omega(N)} \sum_{[f] \in B_N(e)} F(f) \,.
\end{align}
Here, $\Omega(N)$ the volume of $B_N(e)$, defined so that for all $N$, the constant function evaluates to one:
\begin{equation}
    1 = \Delta_N(1) =  \frac{1}{\Omega(N)} \sum_{f \in B_N(e)} 1 \,. \label{eqn:normalized}
\end{equation}
$\Delta_N$ is the distribution which uniformly averages functions $F$ over the ball $B_N(e)$. Other weights with respect to $[f]$ would correspond to different measures. This is well defined for any fixed $N$. However, our goal is to define a sensible measure on \emph{all} of $\text{Mod}(g,n)$. $N$ is essentially an IR regulator for this average, so heuristically, the measure we want to define is obtained from removing this regulator, taking the large $N$ limit
\begin{align}
    \langle F \rangle_{naive} := \lim_{N \to \infty} \Delta_N(F) \,. \label{eqn:largeNlimit}
\end{align}

However, we immediately run into a problem. The mapping class group $\text{Mod}(g,n)$ is not a so-called amenable group \cite{MCGnonamenable},\footnote{One definition of an amenable group is essentially that $\langle F \rangle_{naive}$ exists for bounded functions.} and therefore $\langle F \rangle_{naive}$ does not converge to a well defined distribution for many functions $F$. We explain this and give an example of a bounded function $F_\pm$ with an ill-defined large-$N$ limit in Appendix~\ref{sec:classfunctions}. However, we do not need our distribution $\langle F \rangle$ to exist for \emph{all} functions, but only the class of functions that will actually arise in our computations. This is ultimately what will make our averages over the mapping class group well defined.

As we explained in Sec.~\ref{sec:linkstates}, a fibered link state always takes the form 
\begin{align}
    \ket{M(\La^n)} &= \mathcal{S}\sum_J \tr(K_f[J]) \ket{J}\,.
\end{align}
Because of this, the class of functions we care about will essentially be products of traces $\tr(K_f[J])$ for various $J$. Because $K_f[J]$ is unitary for any $J$, its eigenvalues are all pure phases. This implies that the trace of $K_f[J]$ is bounded above by
\begin{align}
    |\tr(K_f[J])| \leq \dim(\Ha(g,J))\,. \label{eqn:trKfbounded}
\end{align}
It is useful to define a normalized trace
\begin{align}
    \widetilde{\rho}_J(f) = \frac{\tr(K_f[J])}{\dim(\Ha(g,J))}\,. \label{eqn:rhotildedef}
\end{align}
This normalized trace $\widetilde{\rho}_J(f)$ can also be thought of as the eigenvalues of the monodromy operator $\mathcal{P}(f)$ defined in \cite{fiberedlinks}. In terms of the normalized traces, \eqref{eqn:trKfbounded} now reads
\begin{align}
    |\widetilde{\rho}_J(f)| \leq 1 \,. \label{eqn:rhobounded}
\end{align}
This also implies that products of these functions are bounded above:
\begin{align}
    |\widetilde{\rho}_{J_1} \cdots \widetilde{\rho}_{J_N}(f)| \leq 1 \,. \label{eqn:prodofrhobounded}
\end{align}
Each $\widetilde{\rho}_J$ is a function of the mapping class group. There are finitely many such functions, one for each choice of representations $J = j_1, \cdots, j_n$ for the link components. 

Let $\mathcal{F}$ be the subring of functions generated by the functions $\widetilde{\rho}_J$, as well as the constant function $1$. In other words, $\mathcal{F}$ is the set of functions that take the form
\begin{align}
    \mathcal{F} = \text{span}\left\{1,\, \widetilde{\rho}_{J_1}(f)\,,\, \widetilde{\rho}_{J_1} \widetilde{\rho}_{J_2}(f)\,,\, \widetilde{\rho}_{J_1} \widetilde{\rho}_{J_2}\widetilde{\rho}_{J_3}(f)\,, \cdots \right\}\,, \label{eqn:Ffindimbasis}
\end{align}
for any finite sequence $\{J_1, \cdots , J_N\}$ of representations. Functions in $\mathcal{F}$ are automatically bounded functions because each generator $\widetilde{\rho}_J(f)$ is bounded. It is precisely this class of functions that we will be interested in averaging over the mapping class group. The fact that this ring of functions is finitely generated and bounded will be the key to making sure these averages are well defined, despite the nonamenability of the mapping class group.

Because the class of functions $\mathcal{F}$ defined above is finitely generated, we can view it abstractly as a countably infinite dimensional vector space, with a basis given by \eqref{eqn:Ffindimbasis}. The dual space $\mathcal{F}^*$ is then well behaved enough that for any fixed choice of functions $C(J_1, \cdots, J_N)$ for each $N$, we can define a distribution $\langle \cdots \rangle_{\mathcal{F}}$ such that
\begin{align}
    \langle \widetilde{\rho}_{J_1} \cdots \widetilde{\rho}_{J_N} \rangle_{\mathcal{F}} := C(J_1, \cdots, J_N) \,.
\end{align}
By linearity, this definition of $\langle \cdots \rangle_{\mathcal{F}}$ extends to arbitrary functions $F \in \mathcal{F}$. Furthermore, if the complex numbers $C(J_1, \cdots, J_N)$ defining $\langle \cdots \rangle_{\mathcal{F}}$ are bounded, we can use the Hahn-Banach theorem \cite{kolmogorov1999elements} to show that $\langle \cdots \rangle_{\mathcal{F}}$ extends to a distribution on the entire mapping class group $\langle \cdots \rangle_{\text{Mod}(g,n)}$. This distribution is defined such that if $F \in \mathcal{F}$, then
\begin{align}
    \langle F \rangle_{\text{Mod}(g,n)} = \langle F \rangle_{\mathcal{F}}\,. \label{eqn:hahnbanach}
\end{align}
There are many choices of this extension $\langle \cdots \rangle_{\text{Mod}(g,n)}$, and these choices will generally disagree for functions $F' \not\in \mathcal{F}$. However, this disagreement is irrelevant for our purposes, so we can define $\langle \cdots\rangle_{\text{Mod}(g,n)}$ to be \emph{any} such distribution. We will drop the subscript, and will now refer to any average on the mapping class group satisfying \eqref{eqn:hahnbanach}  by $\langle F \rangle$.\footnote{One might worry that $K_f[J]$ is only a projective representation, so the additional phase factor may affect the averages we take. But this extra phase is $J$ independent \cite{Witten:1988hf} and therefore cancels out of $\Psi$ itself, and won't affect the averaging we do.}

In this paper, we will be interested in a particular distribution on $\text{Mod}(g,n)$, defined to be the distribution on $\text{Mod}(g,n)$ such that 
\begin{align}
     C(J_1, \cdots, J_N) = \langle  \widetilde{\rho}_{J_1} \cdots \widetilde{\rho}_{J_N} \rangle_{naive}\,,\label{eqn:preprovenaive}
\end{align}
where $\langle F \rangle_{naive}$ is defined in \eqref{eqn:largeNlimit}. 


Below, we will derive an expression for $\langle  \widetilde{\rho}_{J_1} \cdots \widetilde{\rho}_{J_N} \rangle_{naive}$ using group theory techniques. Technically, we will prove this by averaging the  un-normalized traces $\tr(K_f[J])$ instead. For polynomial functions of $\widetilde{\rho}_J$, of which \eqref{eqn:preprovenaive} is an example, this is equivalent. This is because we can absorb the relative factor of $\dim(\Ha(g,J))$ into the scalars multiplying each coefficient of each term of the polynomial. 

By \eqref{eqn:prodofrhobounded}, this choice of $C(J_1, \cdots, J_N)$ is bounded above for any $N$. Thus, the Hahn-Banach theorem applies, and the distribution $\langle F \rangle_{\text{Mod}(g,n)}$ defined by this procedure is well defined. By construction, averages of functions $F \in \mathcal{F}$ agree with the naive average \eqref{eqn:largeNlimit}:
\begin{align}
    \langle F \rangle = \langle F \rangle_{naive} \,.
\end{align}
Thus, $ \langle F \rangle$ can be thought of as a regulated and renormalized version of \eqref{eqn:largeNlimit}. We can therefore use the naive average $\langle F \rangle_{naive}$ to compute these averages, with the implicit understanding that the averages are actually to be computed using the well defined distribution $\langle F \rangle $ via the above algorithm. This is the procedure we will adopt for the rest of this paper. Because the true distribution is designed to reproduce this formal average, there is no ambiguity in doing so. 

The notation $\langle F \rangle$ is not always a convenient one for computations. We therefore adopt the additional notation
\begin{align}
    \langle F \rangle := \sum_{[f]} F(f) \label{eqn:altavgnotation}
\end{align}
which highlights that $\langle F \rangle$ is to be thought of as a uniform average over $\text{Mod}(g,n)$. Formally, the important feature of this sum is that for all $\psi \in \text{Mod}(g,n)$,
\begin{equation}
    \sum_{[f]} = \sum_{[f \cdot \psi]}\,.
\end{equation}
We mean by this that 
\begin{align}
    \sum_{[f]} F(f \cdot \psi) = \sum_{[f \cdot \psi]} F(f \cdot \psi) = \sum_{[f]} F(f) \,. \label{eqn:shiftingargument}
\end{align}
To see the intuitive reason for \eqref{eqn:shiftingargument}, decompose $[\psi] = t_1 \cdots t_M$ into a minimal product of generators defined by our choice of gates $\mathcal{T}$. Then for any $[f] \in B_{N}(e)$, we know that $[f] \cdot \psi \in B_{N+M}(e)$. But in the limit that $N \to \infty$, this implies that for any fixed $[f]$ and $[\psi]$, we have that $[f \cdot \psi]$ is in the support of $\sum_{[f]}$. Furthermore, because $\text{Mod}(g,n)$ is discrete, the fact that $\sum_{[f]}$ weighs all the points of $\text{Mod}(g,n)$ equally means that this is support containment is sufficient to conclude \eqref{eqn:shiftingargument}. Note that this also shows that $\langle \cdots \rangle$ is independent of the choice of $\mathcal{T}$, even if $\Delta_N$ does for every $N$. Intuitively, $\mathcal{T}$ may control the ``rate'' that $\Delta_N$ approaches $\sum_{[f]}$ along different directions within $\text{Mod}(g,n)$, but the limiting measure is universal. 

An alternative perspective is that \eqref{eqn:shiftingargument} is the \emph{definition} of the formal average $\langle F \rangle_{naive}$ we use to define the ``actual'' probability distribution $\langle F \rangle$ in the above procedure. Either way, we will use \eqref{eqn:shiftingargument} to compute averages over $\text{Mod}(g,n)$, which agrees with the renormalized measure $\langle F \rangle$ as described above.

\subsection{Random links are hyperbolic}\label{sec:hyperbolic}

We call a monodromy $[f]$ hyperbolic (pseudo-Anosov) if its associated link complement can be endowed with a complete hyperbolic metric. Let $\Theta(f)$ be the indicator function on hyperbolic monodromies. In other words, $\Theta(f) = 1$ if $[f]$ is hyperbolic, otherwise $\Theta(f)=0$. Then
\begin{equation}
    P(N) := \Delta_N(\Theta) = \frac{1}{\Omega(N)} \sum_{f \in B_N(e)} \Theta(f)
\end{equation}
is the probability that a randomly chosen monodromy, contained within $B_N(e)$, is hyperbolic. One can show \cite{Maher_2012, rivin2007walksgroupscountingreducible, atalan2008numberpseudoanosovelementsmapping} that there is a constant $\alpha$ such that 
\begin{align}
    P(N) = 1- \mathcal{O}(e^{-\alpha N})\,.
\end{align}
In words, it is exponentially likely that a random monodromy (with respect to $\langle \cdots \rangle$) is hyperbolic. Equivalently, for a fixed genus and number of components, a randomly chosen link is exponentially likely to be hyperbolic. Thus, in the $N\to\infty$ limit we will take, almost all of the link states in this average will be hyperbolic. In other words, because the set of nonhyperbolic links are measure zero (with respect to $\langle \cdots \rangle$) in the moduli space of fibered links, averaging bounded functions over the moduli space of hyperbolic fibered links is equivalent to averaging over all fibered links. Of course, this statement depends on the measure $\langle \cdots \rangle$ we chose for the link monodromy. The (IR regulated) flat measure over monodromy is natural enough that we do not question this choice further in this paper, and leave the sensitivity of our results to this choice for future work. 

\section{Random  un-normalized link states} \label{sec:unnormalizedlinks}

To extract the universal features of a typical link state wave function, we average the link state $\Psi$ over monodromies $[f]$ with respect to the measure $\langle \cdots \rangle$ constructed in Sec.~\ref{sec:measures}. As in \eqref{eqn:altavgnotation}, we denote this average by 
\begin{align}
    \langle F \rangle = \sum_{[f]} F(f) \,.
\end{align}
Therefore, the average over  un-normalized link states \eqref{eqn:matrixoflinkstate} is given by
\begin{align}
     \left\langle \Psi\right\rangle = \sum_{JL} \langle \Psi_{JL} \rangle \mathcal{S}\ketbra{J}{L}\mathcal{S}^\dagger\,. \label{eqn:plugginginaveragedcoefficient}
\end{align}
Thus, the quantity we need to compute is the averaged coefficient
\begin{align}
     \langle \Psi_{JL} \rangle = \sum_{[f]} \tr(K_f[J])\tr(K_f[L])^*\,.
\end{align}
Before computing this average, we first insert a complete set of states to compute each trace:
\begin{align}
      \Psi_{JL} &= \sum_{\chi, \xi} \bra{\chi} K_f[J]\ket{\chi}\bra{\xi}K_f[L]^\dagger\ket{\xi} \,.
\end{align}
We then define the operator
\begin{align}
    \widetilde{\mathcal{O}}_{\chi\xi} = K_f[J]\ket{\chi}\bra{\xi}K_f[L]^\dagger \,, \label{eqn:Otildedef}
\end{align}
so that the matrix element $\Psi_{JL}$ can be computed using $\widetilde{\mathcal{O}}_{\chi\xi}$ as
\begin{align}
     \Psi_{JL}  &= \sum_{\chi,\xi} \bra{\chi}   \widetilde{\mathcal{O}}_{\chi\xi} \ket{\xi} \,. \label{eqn:PsiJLchiJchiL}
\end{align}
The reason for doing this is because the average over monodromy will lead this operator to have a universal form. In fact, that is the next step we will take: average over monodromy with respect to the measure $\langle \cdots \rangle$. Denoting the average of $ \widetilde{\mathcal{O}}_{\chi\xi}$ by
\begin{align}
     \mathcal{O}_{\chi\xi} =  \langle\widetilde{\mathcal{O}}_{\chi\xi}\rangle \,, \label{eqn:avgOdef}
\end{align}
the average density matrix element is given by  
\begin{align}
    \langle \Psi_{JL}\rangle &= \sum_{\chi,\xi} \bra{\chi}  \mathcal{O}_{\chi\xi}\ket{\xi} \,. \label{eqn:avgPsiviaO}
\end{align}

\subsection{Intertwiners} \label{sec:intertwiners}

We will now determine the averaged operator $\mathcal{O}_{\chi\xi}$. We will do so using a group theory argument. The crucial fact will be that both $K_f[J]$ and $K_f[L]$ both lie in (possibly distinct) unitary representations of $\text{Mod}(g,n)$. 
Thus, we can leverage representation theory to understand the structure of $\mathcal{O}_{\chi\xi}$, which is a map between (possibly distinct) unitary representations of the mapping class group. The averaged operator $\mathcal{O}_{\chi\xi}$ has a special property, which we will now demonstrate:
\begin{equation}
    \mathcal{O}_{\chi\xi} K_f[L] = K_f[J]\mathcal{O}_{\chi\xi} \,. \label{eqn:Oisintertwiner}
\end{equation}

Any operator which satisfies this condition is called an intertwiner. One can think of an intertwiner as a map which interpolates between different representations of a group.
To see that $\mathcal{O}_{\chi\xi}$ is an intertwiner, we expand its definition and calculate 
\begin{align}
    \mathcal{O}_{\chi\xi} K_g[L] &= \sum_{[f]} K_f[J] \ketbra{\chi}{\xi}  K_f[L]^\dagger K_g[L]
    \\&=\sum_{[f]} K_f[J] \ketbra{\chi}{\xi}K_{g^{-1}f}[L]^\dagger
    \\&=\sum_{[f]} K_{gf}[J] \ketbra{\chi}{\xi}  K_f[L]^\dagger
    \\&=K_g[J]\sum_{[f]} K_f[J] \ketbra{\chi}{\xi}  K_f[L]^\dagger
    \\&=K_g[J] \mathcal{O}_{\chi\xi}
\end{align}
where we used \eqref{eqn:shiftingargument} in going between the second and third lines. Again, the fact that this holds is essentially the \emph{definition} of the distribution $\langle \cdots \rangle$, as explained in Sec.~\ref{sec:measures}. We also used the fact that $K_f$ is a representation in the second and fourth lines.

Thus, we have demonstrated that $\mathcal{O}_{\chi\xi}$ is an intertwiner between the unitary representations $\Ha(g,J)$ and $\Ha(g,L)$. The technical notation for this space of maps is $\text{Hom}_{\text{Mod}(g,n)}(J,L)$, but we will use the shorthand $\mathcal{I}_{JL}$ instead. 
An upgraded form of Schur's lemma\footnote{To see the connection to Schur's lemma, if $K_g[J]=K_g[L]$, this would say that $\mathcal{O}_{\chi\xi}$ commuted with the group action. This restricts $\mathcal{O}_{\chi\xi} $ to be proportional to the identity. One way to phrase the reason for this is because the space of intertwiners from an irreducible representation to itself is one dimensional, with a basis vector given by the identity map.} \cite{hall2000elementaryintroductiongroupsrepresentations} says that the set of intertwiners forms a vector space, and is finite dimensional.\footnote{One might worry that this Schur's lemma argument doesn't apply because $\text{Mod}(g,n)$ is not compact, but because $\mathcal{O}_{\chi\xi}$ is a map between finite dimensional representations, we don't need to worry about that here.}
We can give this finite dimensional vector space of intertwiners an orthonormal basis, and denote this basis by
\begin{align}
    \mathcal{I}_{JL} = \text{span}\{I^\alpha_{JL}\} \,.
\end{align}
This basis is normalized so that
\begin{align}
    \tr((I^\alpha_{JL})^\dagger I^\beta_{JL})  =\delta_{\alpha\beta}\,.\label{eqn:intertwinernormalized}
\end{align}
Because $\mathcal{O}_{\chi\xi}$ is an intertwiner, we can expand it in this orthonormal basis with some coefficients $c^\alpha_{\chi\xi}$, so that
\begin{align}
    \mathcal{O}_{\chi\xi} = \sum_\alpha c^\alpha_{\chi\xi} I^\alpha_{JL} \,.
\end{align}
We can determine these coefficients by the inner product
\begin{align}
    c^\alpha_{\chi\xi} &=\tr((I^\alpha_{JL})^\dagger \mathcal{O}_{\chi\xi} )
    \\&= \bra{\xi} \left[ \sum_{[f]} K_f[L](I^\alpha_{JL})^\dagger K_f[J]^\dagger \right]\ket{\chi} \,.
\end{align}
We have obtained this expression by plugging in the definition of $\mathcal{O}_{\chi\xi}$ via \eqref{eqn:Otildedef} and \eqref{eqn:avgOdef}. Because $I^\alpha_{JL}$ is an intertwiner, we can simplify this expression by commuting $K_f[L]$ past $I^\alpha_{LJ}$ to see that
\begin{align}
    c^\alpha_{\chi\xi} &= \bra{\xi} \sum_{[f]} (I^\alpha_{JL})^\dagger K_f[J]K_f[J]^\dagger \ket{\chi}
    \\&= \bra{\xi} (I^\alpha_{JL})^\dagger  \ket{\chi}
\end{align}
because the distribution $\langle \cdots \rangle$ is normalized. Thus,
\begin{align}
    \mathcal{O}_{\chi\xi} = \sum_\alpha \bra{\xi} (I^\alpha_{JL})^\dagger  \ket{\chi} I^\alpha_{JL}\,.
\end{align}
Plugging this into \eqref{eqn:avgPsiviaO}, the averaged density matrix coefficient is given by
\begin{align}
    \langle \Psi_{JL}\rangle &= \sum_{\alpha,\chi,\xi}\bra{\xi} (I^\alpha_{JL})^\dagger  \ket{\chi} \bra{\chi}   I^\alpha_{JL}\ket{\xi}
    \\&= \sum_{\alpha}\tr((I^\alpha_{JL})^\dagger    I^\alpha_{JL})\,.
\end{align}
By \eqref{eqn:intertwinernormalized}, we have now shown that the averaged  un-normalized link state density matrix elements are given by
\begin{align}
    \langle \Psi_{JL}\rangle &= \dim(\mathcal{I}_{JL})\,. \label{eqn:averageunnormalizedcoefficient}
\end{align}

We will now compute the dimensions $\dim(\mathcal{I}_{JL})$ using a version of the Verlinde formula. As discussed above, an intertwiner can be thought of as a map between representations of a group $G$ which preserves the group action in the appropriate way. It accomplishes this task by ``fusing'' together irreducible representations of $G$ into new ones, and maps the resulting fused states to their appropriate representations. Actually, this perspective can be made precise, as chiral vertex operators (the CFT objects subject to fusion rules) can be viewed as a quantum generalization of intertwiners in group theory \cite{Moore:1988qv}. The mathematical structure controlling these objects is called a (modular) fusion category \cite{mfc}.

However, while the Hilbert space $\Ha(g,J)$ is constructed using data from the $G_k$ fusion category, it is important to note that $\mathcal{I}_{JL}$ is the space of intertwiners from $\Ha(g,L)$ to $\Ha(g,J)$ viewed as representations of the \emph{mapping class group}, not as $G_k$ representations. This means that the fusion rules which control $\dim(\mathcal{I}_{JL})$ are \emph{not} the same as the $G_k$ fusion rules. This can be seen most easily when $G = \SU(2)$ and the renormalized level $k+2$ is prime. In this case, it is known that $\Ha(g,J)$ is an irreducible representation of the mapping class group, regardless of the labels $J$ of the marked points \cite{primemcgreps,private}. Schur's lemma then implies that
\begin{align}
    \dim(\mathcal{I}_{JL})_{\text{Mod}(g,n)} = \begin{cases}
        1 & \Ha(g,J) \cong \Ha(g,L)\,, \\
        0 & \text{otherwise}\,. \label{eqn:modgnsu2intertwinersdims}
    \end{cases}
\end{align}

Let us now compare this to the number of intertwiners from $\Ha(g,L)$ to $\Ha(g,J)$ thought of as $G_k$ representations, which we denote by $\dim(\mathcal{I}_{JL})_{G_k}$. 

\paragraph{Warmup with $G_k$ intertwiners:} 
The number of $G_k$--intertwiners between $\Ha(g,L)$ and $\Ha(g,J)$ is the same as the number of ways that the representations $\Ha(g,J)$ and $\Ha(g,L)^* \cong \Ha(g,\overline{L})$\footnote{$\overline{L}$ is the conjugate representation to $L$.} can fuse to the trivial representation. This is essentially the CPT theorem. By \cite{Moore:1988qv}, this space of maps can be computed by pair-of-pants decomposing each Riemann surface and sewing them back together in the appropriate way. Because we are fusing $\Ha(g,J)$ and $\Ha(g,\overline{L})$ into the trivial representation, we can geometrically think of this as introducing an additional pair-of-pants into the simultaneous decomposition of these surfaces, with one puncture connected to each surface, and the final puncture of the additional pair-of-pants in the trivial representation. In other words, we add an additional puncture to each surface $\Sigma_{g,J}$ and $\Sigma_{g,\overline{L}}$ and glue them together, resulting in the single, connected surface $\Sigma' = \Sigma_{2g, J\overline{L}}$.

We can now see that
\begin{align}
    \dim(\mathcal{I}_{JL})_{G_k} = \dim(\Ha(2g, J\overline{L})) \,. \label{eqn:intdimisHadim}
\end{align}
Generically, this does not match \eqref{eqn:modgnsu2intertwinersdims}, demonstrating the need to use the correct fusion rules to compute $\dim(\mathcal{I}_{JL})_{\text{Mod}(g,n)}$, which we will once again denote by $\dim(\mathcal{I}_{JL})$.

For some gauge groups like $\SU(2)$ or $\SU(3)$, there exists partial results \cite{andersen2009reducibilitymappingclassgroup,De_Renzi_2021,romaidis2021mappingclassgrouprepresentations} about the $\text{Mod}(g,n)$ fusion rules that we need to determine $\dim(\mathcal{I}_{JL})$ explicitly. But for more general gauge groups, less is known, and we will leave this general classification problem for future work. Despite this, there are still some generalities that can be exploited. 
The irreducible representations of the mapping class group can still be multiplied via the tensor product, and then decomposed into other irreducible representations under the direct product. This structure is still captured by a fusion category, but we again emphasize it is a distinct fusion category from the one formed by primaries in $G_k$ WZW theory, as described above.

\subsection{Mapping class group representations}

Let $V_c$ be an irreducible unitary representation of the mapping class group. One example of such a representation, for $G_k = \SU(2)_{p-2}$ with $p$ prime, is $\Ha(g,J)$ itself. When $k$ is instead an even integer, $\Ha(g,J)$ generically splits into two irreducible representations \cite{andersen2009reducibilitymappingclassgroup}. In that case, $V_c$ refers to one of these subrepresentations of $\Ha(g,J)$. It is known that these more general representations occur for even levels when $G=\SU(2)$ or arbitrary levels $k\geq 3$ for $G=\SU(3)$ \cite{andersen2009reducibilitymappingclassgroup}. 

We can decompose the tensor product $V_a \otimes V_b$ of mapping class group representations into other representations as
\begin{align}
    V_a \otimes V_b = \oplus_c [N_a]_b^c \, V_c\,, \label{eqn:modgnfusion}
\end{align}
for some coefficients $[N_a]_b^c$, which are diagonalizable when thought of as a matrix in its $b,c$ indices \cite{Etingof2005}.
Again, this is essentially the same mathematical structure as the $G_k$ fusion rules of Wilson lines in Chern-Simons, but the specific labels $a$ and fusion matrices $N_a$ are different.
The associativity of fusing the representations implies that these matrices commute for any $a$.
Therefore, they are simultaneously diagonalizable. We can diagonalize them with a matrix $P$ as
\begin{align}
    [N_{a}]_b^c = \sum_d P_{bd} (\lambda_a)_d (P^\dagger)_{cd} \,, \label{eqn:verlindelike}
\end{align}
where $(\lambda_{a})_d$ labels the eigenvalues of $N_a$, and $P$ is the matrix which diagonalizes $N_a$. $P$ is a unitary matrix \cite{Etingof2005}, and is not the same as the $S$ matrix of the $G_k$ fusion category, but is analogous. In fact, it is known \cite{Etingof2005} that for a fusion category, $(\lambda_a)_d = P_{a d}/P_{0 d}$, so the Verlinde formula still holds for the fusion matrices $N_a$.\footnote{This can be proven by using the fact that $[N_0]_a^b = \delta_a^b$ and that $N_{abc}$ is symmetric.}
We use the notation $P$ instead of $S$ to emphasize the difference between them. The fact that all the $N_a$ commute means that all the fusion matrices have the same $P$. 

Using the fusion matrices $N_a$ for irreducible representations, we can define a more general set of fusion matrices for reducible representations as well. Our convention is that
\begin{align}
    N_{a \oplus b} &= N_a + N_b \,,\label{eqn:Noplus} \\
    N_{a \otimes b} &= N_a N_b \label{eqn:Notimes}\,.
\end{align}
This convention matches nicely with the fusion rule \eqref{eqn:modgnfusion}, as one can check directly. Because all the fusion matrices $N_a$ are simultaneously diagonalizable, the above formula also holds for the eigenvalues $(\lambda_a)_d = P_{ad}/P_{0d}$ as well. Finally, we denote the fusion matrix associated with $\Ha(g,J)$ as $N_J$. When $G_k = \SU(2)_{p-2}$, each $N_J$ is itself a fusion matrix for an irreducible representation, but more generally it is a sum of such fusion matrices. 

To determine a formal expression for $\dim(\mathcal{I}_{JL})$, we can use a similar argument to the one for \eqref{eqn:intdimisHadim}, but with the correct $N_J$ fusion matrices instead of the $G_k$ fusion matrices. The number of intertwiners from $\Ha(g,L)$ to $\Ha(g,J)$ is the number of times the trivial representation appears in the fusion of $\Ha(g,J)$ and $\Ha(g,\overline{L})$. From \eqref{eqn:modgnfusion}, this is computed by the matrix element
\begin{align}
    \dim(\mathcal{I}_{JL}) = [N_J N_{\overline{L}}]^0_0 \,. \label{eqn:modgnintertwinerdims}
\end{align}
Using the Verlinde formula \eqref{eqn:verlindelike}, and defining $p_c = P_{0c} (P^\dagger)_{c0} = |P_{0c}|^2$ (no sum on $c$),
\begin{align}
    \dim(\mathcal{I}_{JL}) = \sum_c p_c (\lambda_J)_c (\lambda_{\overline{L}})_c  \,.
\end{align}
Note that $\sum_c p_c = 1$, so we can think about it as a probability distribution over representations $c$. We will now plug this expression into \eqref{eqn:average un-normalizedcoefficient}. In doing so, it is convenient to define the family of states
\begin{align}
    \ket{\lambda_c} = \frac{1}{\Omega(c)} \sum_J (\lambda_J)_c \mathcal{S}\ket{J} \,,\label{eqn:lambdaket}
\end{align}
where $\Omega(c)$ is a normalization constant defined by
\begin{align}
    \Omega(c) = \sqrt{\sum_J (\lambda_J)_c (\lambda_{\overline{J}})_c} \,. \label{eqn:Omegac}
\end{align}
Note that this is always a finite sum, and therefore these states are well defined.

With this definition, the averaged  un-normalized link state can be written as
\begin{align}
    \langle \Psi \rangle = \sum_c p_c\, \Omega(c)^2 \ketbra{\lambda_c} \,. \label{eqn:averageunnormalizedlinkstate}
\end{align}
Even without explicitly determining the states $\ket{\lambda_c}$, there is one important observation that we can make. Let $\sigma$ be a permutation. Because $\Ha(g,J)$ and $\Ha(g,\sigma(J))$ are isomorphic representations (the punctures on the Seifert surface can be deformed to be in any order we wish), $\ket{\lambda_c}$ is in the symmetric subspace of $\Ha(T^2)^{\otimes n}$. We will return to the importance of this in Sec.~\ref{sec:entropy}.

A different approach to averaging over monodromy would have been to assume that $K_f[J]$ was Haar random for each $J$. If we made this assumption instead, then essentially the same argument as above holds, except with the data $\{p_c, \ket{\lambda_c}\}$ appropriate for the fusion rules of $\SU(N)$. This is one way to think about the role of random unitaries in random quantum circuits \cite{Fisher_2023}.
However, for most choices of $G_k$, this would likely lead to a different state than our result \eqref{eqn:averageunnormalizedlinkstate}. The reason is essentially that the fusion rules of a group's representations uniquely determine the group \cite{joyal1991introduction}, a correspondence known as Tannaka-Krein duality. Because $\SU(N)$ and $\text{Mod}(g,n)$ are such qualitatively different groups, it seems likely that the difference in their fusion rules would lead to different states appearing in \eqref{eqn:averageunnormalizedlinkstate}. But without more information about the specific fusion rules of quantum group representations of $\text{Mod}(g,n)$, we can not rule it out. One can interpret this difference as coming from the constraints of summing over quantum states with a geometric interpretation, as opposed to arbitrary quantum states.

\subsection{A toy model: The average $G_k$ state}

Because the states $\ket{\lambda_c}$ are not generally known, \eqref{eqn:averageunnormalizedlinkstate} is rather abstract. To illustrate how this procedure works in a concrete example, we consider a similar problem, but treat $\Ha(g,J)$ as a $G_k$ representation, rather than a $\text{Mod}(g,n)$ representation. The benefit of this is that \eqref{eqn:intdimisHadim} can be calculated explicitly, and the general structure of \eqref{eqn:averageunnormalizedlinkstate} still emerges in this case, as \eqref{eqn:averageunnormalizedlinkstate} holds for any fusion category. 

In this toy model, we can use the Verlinde formula 
to show that \cite{fiberedlinks,Moore:1988qv, Verlinde:1988sn}
\begin{align}
    \dim(\Ha(2g,J\overline{L})) &= \sum_\ell (S_{0\ell})^{2-4g-2n} \bra{J} \mathcal{S}\ket{\ell}^{\otimes n}\!\bra{\ell}^{\otimes n} \mathcal{S}^\dagger \ket{L} 
    \,. \label{eqn:Hadimension}
\end{align}
Our notation is that $\mathcal{S} = S^{\otimes m}$ for whatever power of $m$ is appropriate for the Hilbert space that $\mathcal{S}$ acts on, in this case $m=n$. Furthermore, $S_{0\ell} = \bra{0} S \ket{\ell}$ is a particular matrix element of the modular $S$ transformation. Using the analog of \eqref{eqn:Hadimension}, we can now show that a typical  un-normalized link state takes the form
\begin{align}
    \langle \Psi \rangle &= \sum_{JL} \dim(\Ha(2g, J\overline{L})) \mathcal{S}\ketbra{J}{L}\mathcal{S}^\dagger
    \\&= \sum_{JL \ell} (S_{0\ell})^{2-4g-2n} \mathcal{S}\ketbra{J} \mathcal{S}^\dagger
    \ket{\ell}^{\otimes n}\!\!
    \bra{\ell}^{\otimes n} 
    \mathcal{S} \ketbra{L} \mathcal{S}^\dagger
    \\&= \sum_{\ell} (S_{0\ell})^{2-4g-2n}  \ket{\ell}^{\otimes n}\!\!\bra{\ell}^{\otimes n} \,.
\end{align}
In the third line, we used two resolutions of the identity to compute the $J,L$ sums. 

We can now see that the general structure of \eqref{eqn:averageunnormalizedlinkstate} holds in this example if we identify
\begin{align}
    p_\ell &\mapsto S_{0\ell}^2\,, \\
    \ket{\lambda_\ell} &\mapsto \ket{\ell}^{\otimes n} \,,\label{eqn:Gkbasis} \\
    \Omega(\ell) &\mapsto S_{0\ell}^{-2g-n}\,.
\end{align}

In particular, it is interesting to note that $\ket{\ell}^{\otimes n}$ completely factorizes across the boundary tori, so it is unentangled in this basis. It is currently unknown what precise entanglement properties $\ket{\lambda_c}$ has for arbitrary representations of the mapping class group, and would be very interesting to classify these states for arbitrary $G_k$. We leave this for future work. However, when $G_k = \SU(2)_{p-2}$ for $p$ prime, we will be able to determine enough of the entanglement structure of $\ket{\lambda_c}$ to show that at least some of the $\ket{\lambda_c}$ are not GHZ entangled, which we will prove in Sec.~\ref{sec:entropy}.

\section{Random normalized link states} \label{sec:normalizedlinks}

The expression we derived in \eqref{eqn:averageunnormalizedlinkstate} is not fully satisfactory because the link state $\Psi$ was not properly normalized. One could worry that when we average over the monodromy $[f]$, the varying trace of $\Psi$, which is unphysical, could affect the relative contributions of different link states in the average. We do not want states with a larger trace to dominate the average simply because their wave function coefficients have not been normalized. The proper average to take, then is of the normalized state
\begin{align}
   \overline{\Psi} \equiv \frac{\Psi}{\tr(\Psi)} \,. \label{eqn:normalizedlinkstate}
\end{align}
Note that the average of this normalized state is \emph{not} the same as dividing the average state by its trace:
\begin{align}
    \langle \overline{\Psi}\rangle  \neq\frac{ \langle \Psi\rangle}{\langle\tr(\Psi)\rangle} \,.\label{eqn:neednormalizationprocedure}
\end{align}
This follows from the fact that the normalized density matrix $\overline{\Psi} $ is not linear in $\Psi$, but we will also explicitly demonstrate below that they are not equal. 

The fact that $\tr(\Psi)$ is in the denominator of \eqref{eqn:normalizedlinkstate} is a technical challenge, as the technique of intertwiners we explained above requires all the matrix elements of $K_f$ to be in the numerator. To solve this problem, we will first compute the alternate average
\begin{align}
    \left\langle \Psi \tr(\Psi)^m\right\rangle \,, \label{eqn:replicaaaverageone}
\end{align}
where $m$ is an arbitrary positive integer. After computing this average as a function of $m$, we will then analytically continue this expression in $m$ and treat it as a complex variable\footnote{For more details about the justification of this procedure, see e.g. \cite{SFEdwards_1975,infobook,Lewkowycz_2013}.}. Finally, we will take the limit of this analytically continued expression as $m \to -1$. 

To compute this average as a function of $m$, we will use the following trick. Suppose we wanted to compute $\tr(\Psi)^m $. By the trace identity
\begin{align}
    \tr(A)_\Ha\tr(B)_\Ha = \Tr(A \otimes B)_{\Ha^{\otimes 2}}\,,
\end{align}
we can trade this product of traces for the linear expectation
\begin{align}
     \tr(\Psi)^m  =  \Tr(\Psi^{\otimes m}) \,.
\end{align}
This is known as the replica trick. Expanding $\Tr(\Psi^{\otimes m})$, we see that
 
\begin{align}
     \Tr(\Psi^{\otimes m}) 
    = \sum_{J_i, L_i} \left[\prod_{i=1}^m\tr(K_f[J_i])\right]^*\left[\prod_{i=1}^m\tr(K_f[L_i])\right] \Tr( \ketbra{J_1}{L_1} \otimes \cdots \ketbra{J_m}{L_m} ) \label{eqn:tracereplica1}
\end{align}
 
We can use the replica trick again, this time on the coefficients
\begin{align}
    &\left[\prod_{i=1}^m\tr(K_f[J_i])\right]^*\left[\prod_{i=1}^m\tr(K_f[L_i])\right] \nonumber\\ = & \Tr\left[ \bigotimes_{i=1}^m K_f[J_i]^\dagger \right]\Tr\left[\bigotimes_{i=1}^m K_f[L_i]\right]\,.
\end{align}
For readability, we adopt the replica notation
\begin{align}
   \big|\vec{J}\, \big\rangle\big\langle\vec{L}\big| &= \ketbra{J_1}{L_1} \otimes \cdots \ketbra{J_m}{L_m} \,, \label{eqn:replicanotation1}
   \\ K_f[\,\vec{J}\,] &=\bigotimes_{i=1}^m K_f[J_i]\,,\label{eqn:replicanotation2}
   \\ K_f[\,\vec{L}\,] &= \bigotimes_{i=1}^m K_f[L_i]\,.\label{eqn:replicanotation3}
\end{align}
The vector notation $\vec{J},\vec{L}$ is a useful notation to organize the many different punctures that appear in these calculations. For clarity, $\vec{J}$ is a $m$-dimensional vector with components $J_i$. The index $i$ labels which replica surface the punctured $J_i$ correspond to. In turn, each $J_i$ is a tuple $J_i = (j^1_i, \cdots, j^n_i)$ of defects on a single twisting surface. With this notation, the argument of \eqref{eqn:replicaaaverageone} becomes
 
\begin{align}
    \Psi \tr(\Psi^m) 
    = \sum_{J',L',\vec{J}, \vec{L}} \Tr(K_f[J',\vec{J}\,])^* \Tr(K_f[L',\vec{L}\,]) \Tr\left(\big|\vec{J}\, \big\rangle\big\langle\vec{L}\big|\right) S\ketbra{J'}{L'} S^\dagger\label{eqn:argumentofreplicaaverageone}
\end{align}
 
We now average \eqref{eqn:argumentofreplicaaverageone} over $\text{Mod}(g,n)$ with respect to $\langle \cdots \rangle$. The steps are identical to Sec.~\ref{sec:unnormalizedlinks}, but with the representations $J,L$ generalized to the reducible representations $(J',\vec{J}),(L',\vec{L})$. We leave the genus dependence of these representations implicit. This shows that 
 
\begin{align}
     \langle \Psi \tr(\Psi)^m \rangle
    &= \sum_{J',L',\vec{J}, \vec{L}} \dim(\mathcal{I}_{J'\vec{J},L'\vec{L}})\Tr\left(\big|\vec{J}\, \big\rangle\big\langle\vec{L}\big| \right) S\ketbra{J'}{L'}S^\dagger\,,
    \\&= \sum_{J',L',\vec{J}, \vec{L}} \dim(\mathcal{I}_{J'\vec{J},L'\vec{L}})\,\big\langle\vec{J}\,\big|\,\vec{L}\big\rangle \,S\ketbra{J'}{L'}S^\dagger\,,
    \\&= \sum_{J',L',\vec{J}} \dim(\mathcal{I}_{J'\vec{J},L'\vec{J}}) \,S\ketbra{J'}{L'}S^\dagger\,. \label{eqn:replicanormalization}
\end{align}
 
As argued above, we can use \eqref{eqn:modgnintertwinerdims} with the appropriate representations to show that
\begin{align}
    \dim(\mathcal{I}_{J'\vec{J},L'\vec{L}}) &=[N_{J'}N_{\overline{L}'} N_{J_1}N_{\overline{L}_1} \cdots N_{J_m} N_{\overline{L}_m}]^0_0\,,
    \\&= \sum_c p_c (\lambda_{J'})_c (\lambda_{\overline{L}'})_c (\lambda_{J_1})_c(\lambda_{\overline{L}_1})_c \cdots (\lambda_{J_m})_c(\lambda_{\overline{L}_m})_c\,.
    \label{eqn:normalizedstatecoefficient}
\end{align}
We remind the reader that $N_J$ is the fusion matrix of the mapping class group representations (which is distinct from the $G_k$ fusion matrices), $p_c = |P_{0c}|^2$ is a product of matrix elements of the matrix $P$ which diagonalizes the fusion matrices (essentially the $S$ transformation but for the mapping class group) and $(\lambda_{J})_c$ labels the eigenvalues of $N_J$. This formula follows from using the associativity of the fusion rules to fuse each representation $J_i$ one at a time.
We now sum \eqref{eqn:normalizedstatecoefficient} over $\vec{J} = \vec{L}$ to see that
\begin{align}
    \sum_{\vec{J}}\dim(\mathcal{I}_{J'\vec{J},L'\vec{J}}) &= \sum_c p_c (\lambda_{J'})_c (\lambda_{\overline{L}'})_c \left(\sum_J(\lambda_{J})_c(\lambda_{\overline{J}})_c\right)^m\,.
    \\&= \sum_c p_c (\lambda_{J'})_c (\lambda_{\overline{L}'})_c \Omega(c)^{2m}\,.\label{eqn:sumoverJnormalizedcoefficient} 
\end{align}
where we simplified this expression by recalling the definition of $\Omega(c)$ in \eqref{eqn:Omegac}. Thus, the effect of the $\tr(\Psi)^m$ in the average is to introduce the factor of $\Omega(c)^{2m}$ in \eqref{eqn:sumoverJnormalizedcoefficient}. Plugging this into \eqref{eqn:replicanormalization} and using the definition of $\ket{\lambda_c}$ in \eqref{eqn:lambdaket}, we see that 
\begin{align}
    \langle \Psi \tr(\Psi)^m \rangle &= \sum_{c} p_c \,\Omega(c)^{2(m+1)} \ketbra{\lambda_c}\,.
\end{align}
We now analytically continue $m \to -1$. In doing so, $\Omega(c)$ dependence drops out, leaving the result
\begin{align}
    \langle \overline{\Psi}\rangle = \sum_{c}  p_c \ketbra{\lambda_c} \,. \label{eqn:averagenormalizedlinkstate}
\end{align}
This is one of our main results. Indeed, compared to \eqref{eqn:averageunnormalizedlinkstate}, $\langle \overline{\Psi} \rangle$ is properly normalized:
\begin{align}
    \tr(\langle \overline{\Psi} \rangle) = \sum_c p_c = \sum_c P_{0c} (P^\dagger)_{c0} = 1\,, \label{eqn:itisnormalized}
\end{align}                
where we have used the fact that each $\ket{\lambda_c}$ is normalized. Note that $\langle \overline{\Psi}\rangle $ being normalized does not assume that $\braket{\lambda_a}{\lambda_b} = \delta_{ab}$. Furthermore, one can explicitly use \eqref{eqn:averageunnormalizedlinkstate} and \eqref{eqn:averagenormalizedlinkstate} to compare $\langle \overline{\Psi} \rangle$ and $\langle \Psi \rangle / \langle \tr(\Psi) \rangle$ and see they are not the same,\footnote{The exception is if $\Omega(c)$ is independent of $c$, which is false e.g. for all non-abelian $G_k$ fusion categories.} verifying \eqref{eqn:neednormalizationprocedure} and justifying our more elaborate procedure for averaging over normalized link states.

The implications of \eqref{eqn:averagenormalizedlinkstate} can be understood as follows. Let $\mathcal{O}$ be a state-independent observable acting on the Hilbert space $\Ha(T^2)^{\otimes n}$ of $n$-component link states in Chern-Simons theory (see Sec.~\ref{sec:linkstates} for more details). Because $\mathcal{O}$ is state independent, the expectation values of $\mathcal{O}$ computed using $\langle \overline{\Psi}\rangle$ are approximately equal to the average value of this expectation value, averaged over all possible link states:
\begin{align}
    \left\langle \tr(\mathcal{O} \overline{\Psi})\right\rangle  =  \tr(\mathcal{O}  \langle\overline{\Psi}\rangle) \,. \label{eqn:observableaverage}
\end{align}
Thus, \eqref{eqn:averagenormalizedlinkstate} captures the average behavior of boundary observables in fibered link states of Chern-Simons theory. 



Finally, we can use this result to easily compute $\langle \overline{\Psi}^{\otimes m}\rangle$. Using the same techniques as above, we can use the result for $\langle \Psi^{\otimes m} \rangle$ with the substitution $(\lambda_J)_c \to \frac{(\lambda_J)_c}{\Omega(c)}$ to compute $ \langle \overline{\Psi}^{\otimes m}\rangle$. Explicitly,
\begin{align}
    \langle \overline{\Psi}^{\otimes m}\rangle &= \sum_{\vec{J},\vec{L}}\sum_c p_c \left[\prod_i \frac{(\lambda_{J_i})_c}{\Omega(c)}\frac{(\lambda_{\overline{L}_i})_c}{\Omega(c)} \right] \mathcal{S}\big|\vec{J}\, \big\rangle\big\langle\vec{L}\big|\mathcal{S}^\dagger \,,
    \\&= \sum_c p_c \, \ketbra{\lambda_c}^{\otimes m}\,. \label{eqn:replicatedaveragestate}
\end{align}

Thus, having multiple tensor factors of $\overline{\Psi}$ in the average leads to more copies of $\ketbra{\lambda_c}$, but with the same coefficients $p_c$. This emphasizes the viewpoint that $\langle \overline{\Psi}\rangle$ should be thought of as a classical mixture of the states $\{\ket{\lambda_c}\}$, with the distribution over states being given by $p_c$. This perspective even holds with an arbitrary number of replicas. 

\section{The entanglement of random link states} \label{sec:entropy}

Let $I(\overline{\Psi})$ be any quantum information measure which is analytic in the matrix elements of $\overline{\Psi}$. Examples include the (smooth) entanglement entropy of various subregions, R\'enyi entropies, reflected entropy, and so on. We will now show that
\begin{align}
    \langle I(\overline{\Psi}) \rangle = \sum_c p_c\, I(\ketbra{\lambda_c})\,. \label{eqn:avginfoquantitiy}
\end{align}
In words, the value of the information measure $I(\overline{\Psi})$, averaged over all fibered link states, can be computed by averaging $I(\ketbra{\lambda_c})$ over the distribution $p_c$, both of which are universal and determined by the fusion rules of quantum representations of the mapping class group.
Before proving \eqref{eqn:avginfoquantitiy}, we will briefly discuss an application.

\subsection{Typical hyperbolic link states are not GHZ}

As a concrete application of this formula, let the information measure $I$ be 0 on any quantum state with purely GHZ-like entanglement, and nonzero otherwise. Then $\langle I(\overline{\Psi}) \rangle$ will be nonzero if and only if $I(\ketbra{\lambda_c})\neq 0$ for some $c$. Thus, because hyperbolic links are generic in the fibered link moduli space, hyperbolic link states will have non-GHZ entanglement if and only if at least one state $\ket{\lambda_c}$ does (to leading order in the complexity of a link's monodromy).

To begin, recall that we showed above that $\ket{\lambda_c}$ is always an element of the symmetric subspace of $\Ha(T^2)^{\otimes n}$. Let $d = \dim(\Ha(T^2))$. For example, if $G_k= \SU(2)_k$, then $d = k+1$. The symmetric subspace of $\Ha(T^2)^{\otimes n}$, which we denote by $V_{sym}$, has dimension
\begin{align}
    \dim(V_{sym}) = {d +n-1 \choose n} \approx d^n\,.
\end{align}
for large $d$. On the other hand, the GHZ subspace of $V_{sym}$ is $d$ dimensional, because up to a change of basis, any symmetric GHZ state will take the form
\begin{align}
    \ket{GHZ,sym} = \sum_j \sqrt{p_j} \ket{j\cdots j}\,.
\end{align}
Note that this includes all unentangled states. 
Therefore, if at least $d+1$ of the states $\ket{\lambda_c}$ are linearly independent, then not only will the typical link state be entangled, we can guarantee that it will contain non-GHZ entanglement.

For concreteness, we will now prove that this holds for $\SU(2)_{p-2}$ Chern-Simons theory with $p$ prime, and then comment on the more general case. To begin, let us analyze the structure of $\ket{\lambda_c}$ in more detail. Recall that when $G_k = \SU(2)_{p-2}$, each representation $\Ha(g,J)$ is irreducible as a representation of the mapping class group. Thus, the matrix element $P_{cJ}$ can be thought of as 
\begin{align}
    P_{cJ} = \bra{c} P \ket{(g,J)}
\end{align}
where $\ket{c},\ket{(g,J)}$ are states in the Hilbert space of $\text{Mod}(g,n)$ representations on which $P$ acts. Again, this is \emph{not} the same as the boundary Hilbert space $\Ha(T^2)^{\otimes n}$. There is no tensor product factorization of $\ket{(g,J)}$ across the boundary tori, because each representation is irreducible. Furthermore, to avoid overcounting states, we only consider $J$ to be defined up to permutations in these states, i.e., we view $\ket{(g,J)} \sim \ket{(g,\sigma(J))}$, and project $\ket{J}$ to the symmetric subspace of $\Ha(T^2)^{\otimes n}$.

Because the eigenvalues can be written as $(\lambda_a)_d = P_{a d}/P_{0 d}$, we can rewrite $\ket{\lambda_c}$ in terms of $P$ by
\begin{align}
    \ket{\lambda_c} \propto \sum_J P_{Jc}\, \mathcal{S} \ket{J} = \left(\sum_J \mathcal{S}\ketbra{J}{(g,J)} P \right) \ket{c} \equiv \Pi^\dagger \ket{c}
\end{align}
where we have defined the map $\Pi$ by
\begin{align}
    \Pi^\dagger = \sum_J \mathcal{S}\ketbra{J}{(g,J)} P \,.
\end{align}
Next, we will prove that $\braket{(g,J)}{(g,L)} = 0$ unless $J = L$. 
To see this, note that in a fusion category, it is always the case that the fusion matrix for the trivial representation $N_0$ is the identity: 
\begin{align}
    [N_0]^a_b = \delta^a_b
\end{align}
for any irreps $a,b$. Using the Verlinde formula \eqref{eqn:verlindelike},
\begin{align}
    \delta_J^L = \sum_c P_{Jc} \frac{P_{0c}}{P_{0c}} P^\dagger_{cL} = \bra{(g,J)} P P^\dagger \ket{(g,L)} = \braket{(g,J)}{(g,L)}\,.
\end{align}
We can use this to show that $\Pi$ is an isometry. We show this directly:
\begin{align}
    \Pi\Pi^\dagger &= \sum_{JL} P^\dagger \ketbra{(g,L)}{L} S^\dagger  S\ketbra{J}{(g,J)} P  
    \\&= \sum_J P^\dagger \ketbra{(g,J)} P
\end{align}
which is a projector, and
\begin{align}
    \Pi^\dagger \Pi & = \sum_{JL} S\ketbra{J}{(g,J)} P P^\dagger \ketbra{(g,L)}{L} S^\dagger 
    \\&= \sum_J S\ketbra{J}{J}S^\dagger 
\end{align}
which is the identity on the symmetric subspace. Thus, $\Pi$ is indeed an isometry. Because the state $\ket{c}$ is arbitrary, the possible states $\ket{\lambda_c}$ will span the entire range of $\Pi^\dagger$. To compute how many independent states are in this range, we can use the fact that isometries have eigenvalues $0,1$, and so the trace of $\Pi^\dagger\Pi$ counts the dimension of its range. Explicitly,
\begin{align}
    \dim(\text{range}(\Pi^\dagger)) = \tr(\Pi^\dagger\Pi ) = \sum_J \braket{J} = {d + n - 1 \choose n} 
\end{align}
because we restricted the $J$ sum to the symmetric subspace. But in particular, this implies that there are approximately $d^n - d$ states $\ket{\lambda_c}$ which do not contain GHZ entanglement. Because the genus was arbitrary, this will continue to hold after we average over the genus of the links.  We have therefore proven that with probability 1, a fibered hyperbolic link state is not GHZ-like.
This proves the property of (fibered) hyperbolic links that was conjectured to hold in \cite{Balasubramanian_2018}. 

One reason this result is interesting is it provides a clear connection between multipartite entanglement and the topology of three-manifolds. The Nielsen-Thurston classification \cite{NTclass} says that there are two types of nonreducible link complements: periodic and hyperbolic. Recently, it was shown that all periodic link complements are GHZ entangled \cite{fiberedlinks}. Our result in this paper is complementary: if a Chern-Simons link state contains non-GHZ entanglement, then with probability 1 the link complement that defines it is hyperbolic. In fact, our result holds in any three-dimensional TFT with a chiral RCFT dual. Because topological field theories describe the vacuum sector of gapped QFTs, these results also have interesting implications for vacuum entanglement in gapped QFTs.

Additionally, there are other connections between hyperbolic link complements and Chern-Simons theory that would be interesting to connect to this work. For example, the volume conjecture \cite{murakami2010introductionvolumeconjecture} says that a certain double-scaled limit of the colored Jones polynomial $V_J(\La^n)$ is related to the hyperbolic volume of the manifold that computes it. The Jones polynomial is related to the monodromy by an $S$ transformation
\begin{align}
    V_J(\La^n)^* = \sum_L \mathcal{S}_{JL} \tr(K_f[L])\,,
\end{align}
which was shown in \cite{fiberedlinks}. Because the entanglement structure of a fibered link state is completely determined by $\tr(K_f[L])$, it is plausible that there is a more direct connection between the hyperbolic volume of a three-manifold and some non-GHZ entanglement measure that this result could shed light on. We leave this interesting question for future work.

Finally, we comment on the case of generic $G_k$. In this case, the representations $\Ha(g,J)$ will generally be reducible. However, using the conventions defined in \eqref{eqn:Noplus}, \eqref{eqn:Notimes} for fusion matrices of reducible representations, the same proofs leading to the formula \ref{eqn:averagenormalizedlinkstate} for $\langle \overline{\Psi}\rangle$ still hold. Furthermore, if we assume each mapping class group irrep $V_c$ appears once inside a unique $\Ha(g,J)$, then the same pigeon-hole principle argument given above will still hold. This is because we can still use the unitarity of $P$ to prove that $\braket{(g,J)}{(g,L)} = 0$ for $J\neq L$ in this case: the fact that $\Ha(g,J)$ and $\Ha(g,L)$ split into direct sums of distinct irreps implies that $\ket{(g,J)}, \ket{(g,L)}$ have orthogonal support. It would be interesting to prove for which $G_k$ this uniqueness property holds. It would also be interesting to see if our proof can be adapted to hold in the more general case.

\subsection{Proof of \eqref{eqn:avginfoquantitiy}}

We will now prove \eqref{eqn:avginfoquantitiy}. First, we first note that it is sufficient to prove the special case where $I$ is homogeneous in $\overline{\Psi}$. In other words, we can assume that for some $m \in \N$,
\begin{align}
    I_m( \lambda \overline{\Psi}) = \lambda^m I_m(\overline{\Psi})\,.
\end{align}
Any other analytic $I$ can be constructed as a linear combination of these homogeneous measures, analogous to how any analytic function is a linear combination of monomials $x^m$. Any such $I_m$ can be thought of as descending from a trace
\begin{align}
    I_m(\overline{\Psi}) = \sum_{\mathcal{O}} \alpha(\mathcal{O}) \Tr(\overline{\Psi}^{\otimes m} \mathcal{O})\,, \label{eqn:monomial}
\end{align}
where $\mathcal{O}$ is an operator acting on $\Ha(T^2)^{\otimes m n}$, and $\alpha(\mathcal{O}) \in \C$ is a coefficient. For example, suppose we bipartition the boundary tori into subsets $a$ and $\overline{a}$ as above. Take $\mathcal{O} = \tau_a$ to be the twist operator which cyclically permutes which replica the $d$ punctures $j \in a$ live on, while leaving the $n-d$ punctures $j' \in \overline{a}$ invariant. Then $I_m(\overline{\Psi}) = \tr(\rho_a^m)$. Other $I_m$ correspond to other choices of operators $\mathcal{O}$. The set of information measures spanned by \eqref{eqn:monomial} is complete because we could take $\mathcal{O}$ to be e.g. a projection onto a particular product of $m$ different matrix elements. We can now average $I_m$ and use \eqref{eqn:observableaverage} and \eqref{eqn:replicatedaveragestate} to see that
\begin{align}
    \langle I_m(\overline{\Psi}) \rangle 
    &= \sum_{\mathcal{O}} \alpha(\mathcal{O}) \Tr(\langle \overline{\Psi}^{\otimes m} \rangle \mathcal{O})\,,
    \\&= \sum_c p_c \sum_{\mathcal{O}}\alpha(\mathcal{O})  \Tr(\ketbra{\lambda_c}^{\otimes m} \mathcal{O})\,,
    \\&= \sum_c p_c \, I_m(\ketbra{\lambda_c})\,.
\end{align}
This proves \eqref{eqn:avginfoquantitiy}.

\section{Summary and Discussion} \label{sec:discission}

In this paper, we computed the average quantum-information theoretic behavior of geometric states in Chern-Simons theory. We focused on the case of fibered link states: that is, states prepared by the Chern-Simons path integral on a three-manifold that can be thought of as a twisted circle bundle over a two-surface with boundary. We then defined a (renormalized) uniform average over these geometries. This convention for random link states not only averages over links $\La^n$ in a fixed background manifold $M$, but also sums over possible background manifolds $M$ that the components of $\La^n$ are embedded within, reminiscent of the sum over geometries in the gravitational path integral of three-dimensional gravity \cite{DESER1984220,Coussaert_1995,WITTEN198846,VERLINDE1990652,carlip2023quantumgravity21dimensions}. We can make this analogy more precise as follows. For a fixed number $n$ of boundary tori, let $A_\partial$ denote the boundary values of the Chern-Simons gauge field.
We define a sum over geometries $\mathcal{M}=M(\La^n)$ as
\begin{align}
    \sum_{\mathcal{M}} = \sum_{[f]} \sum_g
\end{align}
where the $[f]$ sum is the average over monodromy we defined in Sec.~\ref{sec:measures}, and the $g$ sum ranges over the genus of the Seifert surfaces of the relevant link. This sum exhausts the moduli space of $n$-component fibered link states. In this spirit, we could define an averaged partition function 
\begin{align}
    \mathcal{Z}[A_\partial] = \langle Z[M] \rangle = \sum_{\mathcal{M}}\int DA \exp(ik I_{CS}[M,A]) \,.
\end{align}

After additionally summing over $g$, the typical state $\langle \overline{\Psi}\rangle$ of \eqref{eqn:averagenormalizedlinkstate} can be thought of as the wave function generated by slicing open the boundary path integral $\mathcal{Z}$, as opposed to $Z$. Because we only averaged over \emph{fibered} links, the gravitational analog of this constraint is not allowing topology change of the wormhole $\Sigma_{g,n}$ connecting the $n$ boundary tori.  We can think of this sum as the leading order term in an expansion of the moduli space of \emph{all} $n$-component link states, organized by the number of times the $S^1$ fibration of \eqref{eqn:fibration} undergoes topology change \cite{morsenovikov}. Any fibration of a given compact three manifold can only undergo topology change a finite number of times, so summing over the number of ``critical points'' where the Seifert surface undergoes topology change would exhaust the moduli space of three manifolds with $n$ toroidal boundaries. 
Because the entanglement of nonfibered link states is always dominated by some fibered link state \cite{fiberedlinks}, we do not expect our results to change by including these topology changing fibrations, but it would be interesting to see if also summing over these geometries changes our results. 

Of course, even in three dimensions where gravity is topological, there are many more degrees of freedom to gravity than just the sum over topologies. For example, in 3D gravity, the modular parameters $\tau_i$ of the $n$ boundary tori will affect the value of the path integral, and e.g. control the phase transition between thermal AdS and BTZ black holes \cite{cmp/1103922135}. Nevertheless, there are many parallels between gravity and summing TFTs over topologies, including a toy model version of the AdS/CFT correspondence \cite{Dymarsky_2025,Banerjee_2022,PhysRevLett.134.151603,Dymarsky:2025agh}. Our work can be understood as an investigation of the entanglement behavior in this toy model. It would be very interesting to understand the relationship of this work to genuine quantum gravity in more detail.

Finally, we note that all results of this paper hold for ``typical'' links, defined with respect to the (renormalized) flat measure over monodromy $\langle \cdots \rangle$. This measure treats all elements of $\text{Mod}(g,n)$ equally: indeed, this is the very feature that led to the universal form for the averages we computed. However, another interesting measure we could define ``typical'' links with respect to would be the flat measure over \emph{conjugacy classes} $[f] \sim [g f g^{-1}]$. This is an interesting choice because links with conjugate monodromies represent homeomorphic link complements \cite{mappingclass}. That being said, link states with conjugate monodromy have distinct quantum states in the boundary Hilbert space, but can only differ by local unitaries \cite{fiberedlinks}. So we could think about such an average as representing a more basis-independent average over link states.


\paragraph{Acknowledgments:} 
It is a great pleasure to thank to Chitraang Murdia for detailed feedback about this work.
We also thank Qingyue Wu, Philip Gressman, and Denis Osin for useful discussions about averaging over the mapping class group. Additionally, we thank Ahmed Barbar for helpful discussions about representations of the mapping class group. It is also a pleasure to thank Vijay Balasubramanian for many helpful conversations throughout this project.
C.C. is supported by the National Science Foundation Graduate Research Fellowship under Grant No.  DGE-2236662. 

\appendix

\section{Nonamenable groups} \label{sec:classfunctions}

In this Appendix, we will explain more of the details behind the measure $\langle F \rangle$ of \eqref{eqn:largeNlimit}. Taken at face value, this limit is only well defined for ``nice enough'' functions $F$.
At first, one might have thought that any bounded function\footnote{A bounded function is a function $F$ with the property that there exists a constant $K$ such that for all $[f] \in \text{Mod}(g,n)$, $F([f]) \leq K$.} would be an acceptable input for $\langle F \rangle$, for it would seem that the average value of $F$ should eventually converge. However, the geometry of $\text{Mod}(g,n)$ with the word metric is exotic enough that this does not always happen.

\begin{figure}
    \centering
 
    \includegraphics{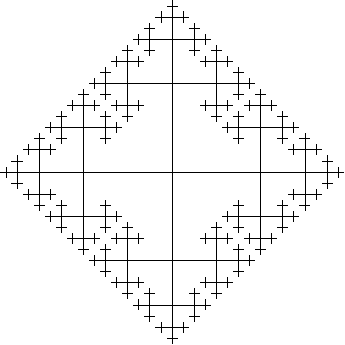}

    \caption{Cayley graph of the prototypical example of a nonamenable group: the free group on 2 generators. The central node represents the identity. An upward move corresponds to right multiplication by $a$, and downward by $a^{-1}$. A rightward move  corresponds to right multiplication by $b$, and downward by $b^{-1}$. The distance of a node from the identity is the number of edges between them. This graph should be continued to an arbitrary number of steps, but we draw $N=5$. In other words, we have drawn $B_5 \subset F_2$. }
    \label{fig:cayley}
\end{figure}

A simpler example of a nonamenable group is $F_2$, the free group on two elements. $F_2$ is the group of finite words such as $aaabaab\cdots$ that can be built out of two letters $a$ and $b$. Together with the word metric, the geometry of $F_2$ is captured by its \emph{Cayley} graph, depicted in Fig.~\ref{fig:cayley}. It is easy to see in this example that the $N$-radius ball has volume $\Omega(N) = 4^N$, which grows faster than more familiar groups such as $\R^d$. This growth rate of $B_N$ is  related to the nonamenability of $F_2$.\footnote{In fact, von Neumann conjectured that any nonamenable group contains $F_2$ as a subgroup. This turns out to be false \cite{Ol2003}, but is true in many examples, including the mapping class group for most $(g,n)$.} The mapping class group $\text{Mod}(g,n)$ contains $F_2$ as a subgroup for $g \geq 2$, so its geometry will share many of the same qualitative features as $F_2$.

\subsection{An example of a nonconvergent function}

Recall our definition
\begin{align}
    \Delta_N(F) = \frac{1}{\Omega(N)} \sum_{f \in B_N(e)} F(f) \,,
\end{align}
where $B_N(e)$ is the $N$-radius ball with respect to the word metric, and $\Omega(N)$ is the volume of $B_N$.
Not all functions $F$ have a well defined $N\to\infty$ limit under this family of distributions.
As an example of such a function, consider
\begin{align}
    F_\pm([f]) = (-1)^{d(e,[f])}\,,
\end{align}
which takes the value $+1$ if $[f]$ is built from an even number of generators in $\mathcal{T}$, and $-1$ if $[f]$ is built from an odd number of generators. We will now show that $\Delta_N(F_\pm)$ does not have a well defined large $N$ limit. First, we note that using the definition of $\Delta_N$, the numerical value of
\begin{align}
    \Delta_N(F_\pm) &= \frac{1}{\Omega(N)} \sum_{f \in B_N(e)} (-1)^{d(e,[f])} 
\end{align}
is well defined for any $N$. Additionally, we can think of $B_N(e)$ as $B_{N-1}(e)$ plus a shell $\partial B_N(e)$ surrounding it. As we are interested in the large $N$ limit, it will be useful to focus on the behavior of the average near this boundary. To do so, we decompose the sum as
\begin{align}
    \Delta_N(F_\pm) =& \frac{1}{\Omega(N)} \sum_{f \in \partial B_N(e)} (-1)^{d(e,[f])} \\&+ \frac{1}{\Omega(N)} \sum_{f \in B_{N-1}(e)} (-1)^{d(e,[f])} \,.
\end{align}
Notice that if $[f] \in \partial B_N(e)$, then by definition $d(e,[f]) = N$, so $F_\pm([f])$ is constant on $\partial B_N(e)$. Furthermore, we can multiply and divide the second term by $\Omega(N-1)$ to see that
\begin{align}
    \Delta_N(F_\pm) &= \frac{\Omega(N) - \Omega(N-1)}{\Omega(N)} (-1)^{N} + \frac{\Omega(N-1)}{\Omega(N)} \Delta_{N-1}(F_\pm) \,.
\end{align}
In the first term, we have used the fact that $|\partial B_N(e)| =\Omega(N) - \Omega(N-1)$. If the large $N$ limit is well defined, then the difference $\Delta_N(F_\pm) - \Delta_{N-1}(F_\pm)$ should vanish as $N \to \infty$. Based on the above calculation, this difference is given by
 
\begin{align}
    |\Delta_N(F_\pm) - \Delta_{N-1}(F_\pm)| &= \left| 
     \frac{\Omega(N) - \Omega(N-1)}{\Omega(N)} (-1)^{N} + \frac{\Omega(N-1) - \Omega(N)}{\Omega(N)} \Delta_{N-1}(F_\pm) 
    \right|
    \\&= \left( 1 - 
     \frac{\Omega(N-1)}{\Omega(N)}
    \right) \left| 
    (-1)^{N} - \Delta_{N-1}(F_\pm) \right| \,.\label{eqn:largeNfail}
\end{align}

The factor $\left| (-1)^{N} - \Delta_{N-1}(F_\pm) \right|$ in this product does not vanish in the large $N$ limit. If it did, then we would have that $\Delta_{N}(F_\pm) \sim (-1)^{N+1}$, a contradiction with \eqref{eqn:largeNfail} because this would imply $|\Delta_N(F_\pm) - \Delta_{N-1}(F_\pm)| \sim 2$. Therefore, for this difference to vanish in the large $N$ limit, the first term $1 - \frac{\Omega(N-1)}{\Omega(N)}$ needs to go to zero as $N \to \infty$. If $\Omega(N) \sim N^d$ scaled as a power law, then we would have
\begin{align}
    1 - \frac{\Omega(N-1)}{\Omega(N)} = 1 - \left(\frac{N-1}{N}\right)^d \approx \frac{d}{N}
\end{align}
and indeed, we would have convergence in the large $N$ limit. This is the reason measures such as $\Delta$ \emph{do} exist for arbitrary bounded functions in e.g. $\R^d$. However, for the mapping class group $\text{Mod}(g,n)$, $\Omega(N)$ scales faster than any power law. To see this, note that any element of $\partial B_N(e)$ can be thought of as an element of $\partial B_{N-1}(e)$ with an additional element from $\mathcal{T}$ applied to it. There are $|\mathcal{T}|$ such possible elements. This implies that to leading order,
\begin{align}
    |\partial B_N(e)| \approx |\mathcal{T}| \cdot |\partial B_{N-1}(e)|\,.
\end{align}
We can iterate this relation, and use that $|\partial B_1(e)| = |\mathcal{T}|$ to see that
\begin{align}
    |\partial B_N(e)| \approx |\mathcal{T}|^N\,.
\end{align}
But $\partial B_N(e) \subset B_N(e)$, and therefore
\begin{align}
    \Omega(N) > |\mathcal{T}|^N\,.
\end{align}
Thus, as long as $|\mathcal{T}| > 1$,
\begin{align}
    1 - \frac{\Omega(N-1)}{\Omega(N)} > 1 - \frac{1}{|\mathcal{T}|} > 0 \,. \label{eqn:volumeratio}
\end{align}
This demonstrates that $\Delta_N(F_\pm)$ does not converge in the large $N$ limit.

Groups with the property that measures defined analogously to $\Delta_N$ converge in the large $N$ limit for bounded functions are called amenable groups. Every compact group is amenable, as is every Abelian group. Furthermore, any semisimple Lie group is amenable. $\text{Mod}(g,n)$, on the other hand, is not amenable, as we demonstrated through the example of $F_\pm$. 
For a more general function $F$, the analog of \eqref{eqn:largeNfail} is
\begin{align}
    |\Delta_N(F) - \Delta_{N-1}(F)| = \left( 1 - 
     \frac{\Omega(N-1)}{\Omega(N)}
    \right) \left| 
    \Delta^\partial_N(F) - \Delta_{N-1}(F_\pm) \right|
\end{align}
where $\Delta^\partial_N$ is the analog of $\Delta_N$, restricted to the boundary $\partial B_N(e)$. Because the first term will never vanish, as was demonstrated in \eqref{eqn:volumeratio}, we can see that to guarantee convergence we must restrict the domain of $\Delta$ to functions such that the \emph{second} factor converges.\footnote{This is even true for amenable groups, if we took the limit via a subsequence $\Delta_{M}$, with $M \sim 2^N$. For amenable groups, however, this class of convergent functions is broader than for nonamenable groups. } In words, we must only allow functions such that do not oscillate too rapidly near the boundary $\partial B_N(e)$.\footnote{This is how we constructed the counterexample $F_\pm$: it was designed to oscillate strongly near the boundary $\partial B_N(e)$ as we vary $N$.} 

\bibliographystyle{hunsrt}
\bibliography{biblio}

\end{document}